\documentstyle[12pt,epsfig]{article}

\def\blacktriangleup{\vrule width0.25pt height0.5pt depth0pt \hskip0pt
\vrule width0.25pt height1pt depth0pt \hskip0pt
\vrule width0.25pt height1.5pt depth0pt \hskip0pt
\vrule width0.25pt height2pt depth0pt \hskip0pt
\vrule width0.25pt height2.5pt depth0pt \hskip0pt
\vrule width0.25pt height3pt depth0pt \hskip0pt
\vrule width0.25pt height3.5pt depth0pt \hskip0pt
\vrule width0.25pt height4pt depth0pt \hskip0pt
\vrule width0.25pt height4.5pt depth0pt \hskip0pt
\vrule width0.25pt height5pt depth0pt \hskip0pt
\vrule width0.25pt height5.5pt depth0pt \hskip0pt
\vrule width0.25pt height6pt depth0pt \hskip0pt
\vrule width0.25pt height6.5pt depth0pt \hskip0pt
\vrule width0.25pt height7pt depth0pt \hskip0pt
\vrule width0.25pt height7.5pt depth0pt \hskip0pt
\vrule width0.25pt height8pt depth0pt \hskip0pt
\vrule width0.25pt height8.5pt depth0pt \hskip0pt
\vrule width0.25pt height9pt depth0pt \hskip0pt
\vrule width0.25pt height8.5pt depth0pt \hskip0pt
\vrule width0.25pt height8pt depth0pt \hskip0pt
\vrule width0.25pt height7.5pt depth0pt \hskip0pt
\vrule width0.25pt height7pt depth0pt \hskip0pt
\vrule width0.25pt height6.5pt depth0pt \hskip0pt
\vrule width0.25pt height6pt depth0pt \hskip0pt
\vrule width0.25pt height5.5pt depth0pt \hskip0pt
\vrule width0.25pt height5pt depth0pt \hskip0pt
\vrule width0.25pt height4.5pt depth0pt \hskip0pt
\vrule width0.25pt height4pt depth0pt \hskip0pt
\vrule width0.25pt height3.5pt depth0pt \hskip0pt
\vrule width0.25pt height3pt depth0pt \hskip0pt
\vrule width0.25pt height2.5pt depth0pt \hskip0pt
\vrule width0.25pt height2pt depth0pt \hskip0pt
\vrule width0.25pt height1.5pt depth0pt \hskip0pt
\vrule width0.25pt height1pt depth0pt \hskip0pt
\vrule width0.25pt height0.5pt depth0pt}

\def\blacktriangledown{\raisebox{1.2ex}{\vrule width0.25pt height0pt depth0.5pt 
\hskip0pt
\vrule width0.25pt height0pt depth1pt \hskip0pt
\vrule width0.25pt height0pt depth1.5pt \hskip0pt
\vrule width0.25pt height0pt depth2pt \hskip0pt
\vrule width0.25pt height0pt depth2.5pt \hskip0pt
\vrule width0.25pt height0pt depth3pt \hskip0pt
\vrule width0.25pt height0pt depth3.5pt \hskip0pt
\vrule width0.25pt height0pt depth4pt \hskip0pt
\vrule width0.25pt height0pt depth4.5pt \hskip0pt
\vrule width0.25pt height0pt depth5pt \hskip0pt
\vrule width0.25pt height0pt depth5.5pt \hskip0pt
\vrule width0.25pt height0pt depth6pt \hskip0pt
\vrule width0.25pt height0pt depth6.5pt \hskip0pt
\vrule width0.25pt height0pt depth7pt \hskip0pt
\vrule width0.25pt height0pt depth7.5pt \hskip0pt
\vrule width0.25pt height0pt depth8pt \hskip0pt
\vrule width0.25pt height0pt depth8.5pt \hskip0pt
\vrule width0.25pt height0pt depth9pt \hskip0pt
\vrule width0.25pt height0pt depth8.5pt \hskip0pt
\vrule width0.25pt height0pt depth8pt \hskip0pt
\vrule width0.25pt height0pt depth7.5pt \hskip0pt
\vrule width0.25pt height0pt depth7pt \hskip0pt
\vrule width0.25pt height0pt depth6.5pt \hskip0pt
\vrule width0.25pt height0pt depth6pt \hskip0pt
\vrule width0.25pt height0pt depth5.5pt \hskip0pt
\vrule width0.25pt height0pt depth5pt \hskip0pt
\vrule width0.25pt height0pt depth4.5pt \hskip0pt
\vrule width0.25pt height0pt depth4pt \hskip0pt
\vrule width0.25pt height0pt depth3.5pt \hskip0pt
\vrule width0.25pt height0pt depth3pt \hskip0pt
\vrule width0.25pt height0pt depth2.5pt \hskip0pt
\vrule width0.25pt height0pt depth2pt \hskip0pt
\vrule width0.25pt height0pt depth1.5pt \hskip0pt
\vrule width0.25pt height0pt depth1pt \hskip0pt
\vrule width0.25pt height0pt depth0.5pt \hskip0pt}}

\newcommand{\nd}{\noindent}

\begin{document}
\begin{center}
{\LARGE On the Multiplicity Distributions of Charged Secondaries in the 
Collisions of Relativistic Nuclei}
\end{center}
\medskip

\begin{center}
{\sc S.M. Esakia$^1$, V.R. Garsevanishvili$^{2,3}$, T.R. Jalagania$^4$}\\
{\sc G.O. Kuratashvili$^1$, Yu.V. Tevzadze$^1$}\\

\bigskip

\begin{tabular}{ll}\small
$^1$&Institute of High Energy Physics, Tbilisi\\
    &State University, 380086 Tbilisi, Rep. of Georgia\\
    &E-mail: tevza@sun20.hepi.edu.ge\smallskip\\
$^2$&Laboratoire de Physique Corpusculaire,\\
    &Universit\'e Blaise Pascal, 63177 Aubi\`ere Cedex France\\
    &E-mail: garse@clrvax.in2p3.fr\smallskip\\
$^3$&Mathematical Institute of the Georgian Academy of Sciences,\\
    &380093 Tbilisi, Rep. of Georgia\\
    &E-mail: garse@imath.acnet.ge\smallskip\\
$^4$&Department of Physics, Tbilisi State University,\\
    &380000 Tbilisi, Rep. of Georgia\smallskip\\
    &Contact author: Yu. V. Tevzadze\\
    &E-mail: tevza@sun20.hepi.edu.ge\\
\end{tabular}
\end{center}

\vfill

\begin{center}
{\bf Abstract}
\end{center}

Multiplicities of charged secondary hadrons in the relativistic
nucleus--nucleus collisions in a wide energy range are analysed on the basis
of partial stimulating emission and cluster cascading models. The experimental
data are obtained by means of the two metre propane bubble chamber of 
JINR (Dubna). The results are compared with the corresponding data for
$pp$ and $\bar pp$-collisions at higher energies. It is shown that the
regularities in $pp$ and $\bar pp$-interactions at high energies and wide
energy range and in nucleus-nucleus interactions at relatively low energies
per nucleon and narrower energy range are similar.
\bigskip


\clearpage

\section{Introduction and Basic Notions}

The interest to the study of multiplicity distributions of charged secondaries
has been increased again after the new powerful accelerators have been
constructed
and the beams of protons, antiprotons, electrons, positrons and heavy ions
 have been obtained (see,e.g. [1-5]).

In the present work multiplicities of charged secondary hadrons in 
${\it pp}$, $\bar{{\it p}}{\it p}$ and nucleus-nucleus collisions 
 in a wide energy range are analysed
on the basis of partial stimulating emission (PSE) and cluster cascading models
(CCM). The manifestation of the negative binomial
distribution (Pascal distribution)
\begin{equation}
P_{n}=\frac{(n+k-1)!}{(k-1)!n!}\left(\frac{<n>}{<n>+k}\right)^{n}
\left(\frac{k}{<n>+k}\right)^{k}
\end{equation}
in the multiplicity distributions of secondaries is interpreted in the 
framework of these models.

In Eq.(1) $<n>=<n_{\pm}>$ is the average multiplicity of all charged secondaries.
Parameter $k$ determines the form of the distribution, e.g. if $k^{-1}=1$ we
get the geometrical distribution, if $k^{-1}=0$ we get the Poisson distribution.

It is useful to establish a recurrance relation between $P_{n}$ and
$P_{n+1}$ [6]. When deriving this relation one starts from the consideration
that the event with multiplicity $(n+1)$ can be expressed by means of $(n+1)$ 
number of events with multiplicity $n$:
\begin {equation}
g(n)=\frac{(n+1)P_{n+1}}{P_{n}} 
\end{equation} 

Inserting Eq.(1) into Eq.(2) one can express $g(n)$ as:
\begin{equation}
g(n)=a+b n,
\end{equation}
where:
\begin{equation}
a=\frac{k<n>}{k+<n>},\ \ \ \ \  b=\frac{<n>}{k+<n>}
\end{equation}
From here one can write:
\begin {equation}
<n>=\frac{a}{1-b},\ \ \ \ \   k=\frac{a}{b}
\end{equation}
 
Dispersion $D=(<n^2>-<n>_2)^{1/2}$ of the distribution (1) and 
parameters $<n>$ and $k$ are related as:
\begin{equation}
k=\frac{<n>^2}{D^2-<n>}
\end{equation}
 
PSE model admits the following interpretation of the distribution (1) and the
relation (3) [6]. The emitted particles are uniformly distributed among $k$
cells. These cells do not correlate and there is no connection between the
particles in different cells. The additional $(n+1)$-th particle can be emitted 
in the initial act of the collision independently of the already existing
$n$ particles. This is reflected by the constant term $\alpha$ in the function
$g(n)$. In this case $b=0$ (i.e. $k^{-1}=0$) and the classical Poisson
distribution is obtained. But such an emission can be intensified as a result
of quantum interference effects. The average effect of this intensification
is expressed by adding the linear term $b n\ (b=const)$ in the 
function $g(n)$. It follows from the Eqs. (3) and (5) that:
\begin{equation}
g(n)=a\left(1+\frac{n}{k}\right)
\end{equation}
 
One can conclude that $n/k$ is the average number of particles
among already existing $n$ particles which promote the creation of a new
$(n+1)$-th particle. So $k^{-1}$ is the relative average fraction of particles 
which stimulate such creation.

On the other hand the multiple production can be interpreted in the
framework of CCM. One assumes here that after the collision of high energy
particles (leptons, hadrons, nuclei) some exited $n$-particle system is
produced which is formed as an $N$-cluster state. Each of these clusters is
formed by the particles which are produced directly or indirectly from one
particle produced in the initial act of the collision. These latter particles
are called the "patriarchs" of the clusters. The "patriarch" which does not
produce secondaries forms one particle cluster itself. It is assumed that
"patriarchs" and consequently clusters are produced independently from 
each other. Therefore for the multiplicity of clusters the Poisson
distribution holds:
\begin{equation}
F(N)\sim \frac{1}{N!}<N>^N
\end{equation}
For the better understanding of the CCM let us give a short review of the
formulae of this model. It is evident that the average number of clusters
$<N>$ is given by the formula:
\begin{equation}
<N>=\frac{<n>}{<n_c>}
\end{equation}
where $<n>$ is the average number of charged hadrons, $<n_c>$ is the average
number of hadrons in the cluster.

Let $F_c(n_c)$ be the distribution of particles inside one cluster. It is
assumed in the CCM that the recurrence relation between $F_c(n_c+1)$ and
$F_c(n_c)$ is of the form:
\begin{equation}
g_c(n_c)=\frac{(n_c+1)F_c(n_c+1)}{F_c(n_c)}=bn_c,\ \ \ b=const,\ \ \ n\geq 1
\end{equation}

The meaning of this relation is that the effect of the creation of $(n_c+1)$-th
particle is proportional to the number of particles already existing in the
cluster in average. Iterating Eq.(10) one gets:
\begin{equation}
F_c(n_c)=F_c(1)\frac{b^{n_c-1}}{n_c}
\end{equation}
where $F_c(1)$ can be found from the normalization condition:
\begin{equation}
\sum_{n_c}F_c(n_c)=1
\end{equation}
From here one gets:
\begin{equation}
F_c(1)=-\frac{b}{\ln{(1-b)}}
\end{equation}
 
One can find the following average values:
\begin{equation}
<n_c>=\sum_{n_c}n_cF_c(n_c)=\frac{F_c(1)}{1-b}
\end{equation}
\begin {equation}
<n_c^2>=\sum_{n_c}n_c^2F_c(n_c)=\frac{<n_c>}{1-b}
\end{equation}
 
Note finally that on the basis of Eqs.(8) and (11) one can write the 
expression for the total multiplicity $n$ in the form of the negative
binomial distribution which is represented in the form [6]:
\begin{equation}
P_n\sim a(a+b)[a+b(n-1)]
\end{equation}
where
\begin{equation}
a=<N>F_c(1)
\end{equation} 

Inserting Eq.(16) into the recurrence relation (2) one gets again Eq.(3). The
identification of the parameters $a$ and $b$ which is given by Eqs.(10) and
(17) completly corresponds to their earlier physical meaning. So the CCM is
compatible with the negative binomial distribution.

\section{Analysis of the Experimental Data} 

The experimental data are obtained on the two-metre propane bubble chamber
(PBC-500) of the Laboratory of High Energies of JINR (Dubna) with tantal targets 
in it which were bombarded by ${\it p,d}$, $He$ and $C$ beams[5]. The data on
${\it pp}$ (ISR and others) and $\bar{{\it p}}{\it p}$ (UA5- Collaboration)
collisions in a wide energy range [1-4,6] are used for comparision.

It was shown in Ref.[4] that in ${\it pp}$ and $\bar{{\it p}}{\it p}$-
collisions $(10\leq \sqrt{s}\leq 900 GeV)$ $k^{-1}$ increases linearly and reaches 
the value 0.31 at $\sqrt{s}=900 GeV$, i.e. every third particle is activly
interacting. The average number of charged particles in the cluster $<n_c>$
reaches the value 4.55, the average number of clusters reaches the value 8. May
be there is even a decrease of the number of clusters as compared to the energy
$\sqrt{s}=540 GeV$ (Table 1). The width of the distribution of charged
particles in the clusters increases with increasing energy much faster than
the corresponding dependence for the total multiplicity. The ratio
$R=<n>/D^2$ decreases 12 times in the range $8.33\leq \sqrt{s}\leq 
900 GeV$.
The same quantity for the clusters $R_c=<n_c>/D^2_c$ decreases 
400 times (Tables 1a and 1b, Fig. 1). Such a decrease is observed up to 
$\sqrt{s}=200 GeV$ and after that $R$ and $R_c$ behave almost in the same way.
After $\sqrt{s}=200 GeV$ a tendency of the narrowing of the multiplicity 
distributions in the cluster and the total multiplicity is observed.

What is the behaviour of $R=f(E)$ and $R_c=f(E)$ for ${\it p}Ta$-collisions
in the range 2-10 GeV? A fast decrease of $R$ and especially of $R_c$ is
observed up to the energy 5 GeV, i.e. the dispersion increases fast (Table 2,
Fig. 2) At the energies higher than 5 GeV a change of the regime is 
observed and probably they behave as constants. The constant regime in
${\it pp}$ and $\bar{{\it p}}{\it p}$-collisions is observed from
$\sqrt{s}=200 GeV$.

Consider the same dependences for ${\it d}Ta, HeTa$ and $CTa$-collisions
(Tables 3,4,5). For ${\it d}Ta$-collisions $R_c(E)$ decreases faster than
$R(E)$. The situation is the same as in ${\it p}Ta$-collisions in the same
energy range. In the case of $HeTa$ and $CTa$-collisions the behaviours of
$R_c(E)$ and $R(E)$ are similar.

It is interesting to note that a similar picture arises, if we consider the
dependence of $R$ and $R_c$ on the atomic weight $A_i$ of the projectile
nucleus. For ${\it p}Ta, {\it d}Ta, HeTa$ and $CTa$-collisions at 4.2 AGeV 
$R$ and $R_c$ decrease with increasing $A_i$. But, if $R$ decreases 19 times,
$R_c$ decreases 50 times. Further, the ratio $R_c/R$ is
approximately equal to 8 for ${\it p}Ta$-collisions and is equal to 1.5 for
$CTa$-collisions at the same energy, i.e. this ratio tends to one with
incresasing $A_i$.

In Ref.[7] the dependence $k^{-1}=f(s)$
for ${\it pp}$ and $\bar{{\it p}}{\it p}$-collisions in the range 
$10\leq \sqrt{s}\leq 900 GeV$ is approximated by the formula:
\begin{equation}
k^{-1}=a_1+b_1\ln \sqrt{s}\ , 
\end{equation}
$k^{-1}>0$ and decreases rather slowly with increasing energy ($b_1\approx 
0.06$).
It has been shown by our analysis that at lower energies, in particular,
at $\sqrt{s} < 8.33 GeV$ (corresponding laboratory energy is 36 GeV) 
$k^{-1} < 0$. At higher energies $k^{-1}$ becomes positive.

Consider now nucleus-nucleus collisions. If by analogy with Eq.(18) we
approximate $k^{-1}$ by the formula:
\begin{equation}
k^{-1}=a_2+b_2\ln E,
\end{equation}
it turnes out that in a narrow energy range (2-5 GeV) a fast increase of
$k^{-1}$ is observed, $b_2 \geq 0.15$ (Tables 3,4,5, Fig. 3). For the energies 
higher than 5 GeV the fast increase of $k^{-1}$ is slowed down. The essential
feature of ${\it p}Ta$-collisions is that at the energy 2.48 GeV the 
parameter $k^{-1}$ becomes negative. Starting from 4.3 GeV (may be earlier)
$k^{-1}$ becomes positive, i.e. there is a similarity with ${\it pp}$-
collisions but at lower energies. For ${\it d}Ta, HeTa,CTa$-collisions parameter 
$k^{-1}$ is positive and increaseing with increasing energy, i.e. the growth
of the atomic weight of the incoming nucleus plays the same role as the growth
of energy in ${\it pp}$-collisions (Tables 1, 3-5,Refs.[1-8]).

Consider the situation with negative $k^{-1}$ in some detail. Note that one
writes nothing about them in Ref.[6], though they are obtained from the
experimental data (Tables 1a,2, Refs.[9,10]). Probably the reason is that
the negative bimomial distribution and PSE and CCM in their strict
formulations assume the positivness of $k^{-1}$. But the stable presence
of negative $k^{-1}$ made us to take a broader view to this problem. First of
all, if we consider the negative binomial distribution in the form of Eq.(16)
and compare it with the Polya-Egenberger distribution (see, e.g. [10])
\begin{equation}
P_n=\left(\frac{<n>}{1+g^2<n>}\right)^n\left(\frac{1}{1+g^2<n>}\right)^
{1/g^2}\frac{1}{n!}\sum_{\alpha=0}^{n-1}(1+\alpha g^2),
\end{equation}
where $g^2=(D^2-<n>)/<n>^2$,it is evident that these two distributions
coincide. The role of the parameter $k^{-1}=b/a$ is played
by $g^2$. But the sign of the parameter $g^2$ in the Eq.(20) is not fixed.
In particular, if $g^2=0$, we get the Poisson distribution, if $g^2>0$,
this distribution is broader than the Poisson one, if $g^2<0$, it is narrower 
than the Poisson one. So parameter $g^2$ can be thought as a measure of 
deviation of the distribution from the Poisson one and if we perform the 
analysis of the experimental data in terms of the Polya-Egenberger distribution, 
parameter $k^{-1}=g^2$ can take positive and negative values as well.
Further, there are some grounds to think that the negative values of $k^{-1}$
are compatible with the assumptions of PSE and CCM. In fact, the second term
in Eq.(3) corresponds to quantum-mechanical interference effects. If this
effects stimulate the creation of $(n+1)$-th particle, then $b>0$ and
hence $k^{-1}=b/a>0$, since $a>0$. But it is natural to
suppose that this effects can make weaker the emission (cupture of the
particle). In this case $\beta<0$ and hence $k^{-1}<0$. So one can conclude
that $k^{-1}<0$ corresponds to the cupture of secondary particles (see
also [5,11]).

Proceed now to consider the energy dependence of the total average
multiplicity of particles $<n>$ and the average multiplicity of particles
in the clusters $<n_c>$ in $A_iTa$-collisions and compare this data with
corresponding results on ${\it pp}$ and $\bar{{\it p}}{\it p}$-collisions. 
In the energy range 7.42-200 GeV in ${\it pp}$ and $\bar{{\it p}}{\it p}$-
collisions $<n>$ and $<n_c>$ increase rather fast, but $<n>$ increases
faster than $<n_c>$. In the range (200-900) GeV the increase of these 
quantuties is slower and both of them increase 1.6 times (Table 1).

Consider now ${\it p}Ta, {\it d}Ta, HeTa$ and $CTa$-collisions. In ${\it p}Ta$- 
collisions the energy range can also be divided into two parts. The first
interval (2-5) GeV and the second one (5-9) GeV. In the first interval we
observe a more rapid increase of the average multiplicities $<n>$ and
$<n_c>$ than in the second one (Table 3, Fig. 4). So we have qualitativly
the same situation as in ${\it pp}$ and $\bar{{\it p}}{\it p}$- collisions,
but the energy range is more narrow and low. In ${\it d}Ta, HeTa$ and 
$CTa$-collisions in the energy range (2-5) GeV the average multiplicities
$<n>$ and $<n_c>$ increase similarly with increasing energy.

It is interesting to trace the dependence of $<n>$ and $<n_c>$ on the atomic 
weight $A_i$ of the incoming nucleus at fixed energy. It is seen from
Tables 3,4,5 and Fig. 4 that at 2.5 GeV $<n_c>$ increases 3 times and at
4.3 GeV - 4 times. The behaviour of $<n>$ is similar. So the increase of the 
atomic weight of the incoming nucleus in $A_iA_t$-collisions plays a similar
role as the increase of the energy. As a result in $A_iA_t$-collisions the
same effect is achieved at relativly low energies as compared to hadron-
hadron collisions. This can be confirmed by the following example: in 
$CTa$-collisions at 2.48 AGeV $<n_c>=2.66$ and at 4.30 AGeV $<n_c>=4.55$. 
The same number of particles is contained in the clusters in 
${\bar pp}$-collisions at $\sqrt{s}= 200 GeV$ and $\sqrt{s}=900 GeV$, 
respectively.

It is interesting to study the dependences of the dispersions on 
$<n>$ and $<n_c>$:
\begin{equation}
D=f(<n>),
\end{equation}
\begin{equation}
D_c=f(<n_c>)
\end{equation}

It is seen from Fig. 5 that at the energies higher than 5 GeV the data for
the dependence (21) for hadron-hadron and nucleus-nucleus collisions lie on
two different curves (curves 1 and 2), but the data for the dependence (22)
lie on the same curve (curve 3).

Let us show in conclusion the multiplicity distribution of charged 
secondaries in ${\it d}Ta$-collisions at 5.18 AGeV and its fit according to
Eq.(1) (Fig. 6) with only normalization parameter,which is approximately
equal to one, $\chi^2/N=1.15$, $N$ is the number of experimental points.

The authors are indebted to the staff of two metre propane bubble chamber of
JINR (Dubna) for supplying the data. They would like to thank Ya. Darbaidze,
E. Khmaladze, N. Kostanashvili, P. Pras, G. Roche, T. Topuria, M. Topuridze
for interesting discussions. On of the authors (V.R.G.) expresses
his deep gratitude to Bernard Michel and Guy Roche for the warm hospitality
at the Laboratoire de Physique Corpusculaire, Universit\'e Blaise Pascal,
Clermont-Ferrand, to V. Kadyshesky, T. Kopaleishvili, H. Leutwyler,
W. R\"uhl for supporting his stay at the L.P.C. and to NATO for
supporting this work.

\newpage

\section{References}

\nd 1. A.Breakstone et al. {\it Phys.Rev}., 1984, v.D30, p.528

\nd 2. G.J.Alner et al. {\it Phys. Lett}., 1984, v.B138,p.304
       R.Ansorge et al. {\it Charge Particle Multiplicity Distributions at 
       200 and 900 GeV c.m.energy},UA5-Collaboration, CERN-EP/88-172

\nd 3. M.Althoff et al. TASSO-Collaboration, {\it Z.Phys}., 1984 v.C22, p.307

\nd 4. G.J.Alner et al. {\it Phys.lett}., 1985, v.B160, p.199

\nd 5. E.O.Abdurakhmanov et al. {\it Sov. J. Nucl. Phys.}, 1978, v.27, p.1020;
       1978, v.28, p.1304
       M.A.Dasaeva et al. {\it Sov. J. Nucl. Phys.}, 1984, v.39, p.846    

\nd 6. A.Giovannini, L.Van Hove. {\it Negative Binomial Distribution 
       in High Energy Hadron Collisions.} CERN-TH 4230/85

\nd 7. G.J.Alner et al. {\it Scaling Vilations in Multiplicity Distributions at 
       200 and 900 GeV.} CERN-EP/85-197; {\it Phys.Lett.}, 1986, v.B167, p.476

\nd 8. V.V.Ammosov et al. {\it Phys.Lett.}, 1972, v.B42, p.519
       D.B.Smith et al. {\it Phys.Rev.Lett.}, v.B42, p.519

\nd 9. N.K.Kutsidi, Yu.V.Tevzadze. {\it Sov. J. Nucl. Phys.},1985, v.41, p.236

\nd 10.C.Vokal, M.Shumbera. {\it JINR-Communications,} 1-82-388, Dubna, 1982

\nd 11.N.S.Grigalashvili et al. {\it Sov. J. Nucl. Phys.}, 1988, v.48, p. 476

\newpage
 \begin{center}
{\sc Table 1a}\\
 Characteristics of the Multiplicity Distributions of Charged Hadrons\\
in {\it pp}-collisions\\
\end{center}
$$\begin{tabular}{|c|c|c|c|c|c|}
\hline
 & \multicolumn{5}{|c|}{Energy in the c.m.s. $\sqrt{s}$, GeV}\\
\cline{2-6}
 & 7.42 & 8.33 & 9.78 & 30.4 & 62.2\\
\hline
$<n>$ & 4.56$\pm$0.04 & 4.78$\pm$0.03 & 5.32$\pm$0.13 & 10.7 & 13.6 \\
\hline
$D$ & 2.09$\pm$0.04 & 2.23$\pm$0.03 & 2.58$\pm$0.05 & 4.59 & 6.01 \\
\hline
$<n>/D^2$ & 1.04$\pm$0.02 & 0.96$\pm$0.04 & 0.80$\pm$0.03 & 0.51 & 0.38 \\
\hline
$k^{-1}$ & -0.009 & 0.008 & 0.05 & 0.09 & 0.12 \\
\hline
$b$ & -0.044 & 0.039 & 0.20 & 0.49 & 0.62  \\
\hline
$<n_c>$ & 0.98 & 1.02 & 1.12 & 1.43 & 1.69 \\
\hline
$D_c$ & - & 0.14 & 0.38 & 0.87 & 1.26 \\
\hline
$<n_c>/D_c^2$ & - & 51 & 7.76 & 1.88 & 1.06 \\
\hline
$<n>/k$ & -0.04 & 0.04 & 0.25 & 0.97 & 1.66 \\
\hline
$<N>$ & 4.66 & 4.69 & 4.75 & 7.48 & 8.05 \\
\hline
\end{tabular}$$
\bigskip

 \begin{center}
{\sc Table 1b}\\
 Characteristics of the Multiplicity Distributions of Charged Hadrons\\
in $\bar{{\it p}}{\it p}$-collisions\\
\end{center}
$$\begin{tabular}{|c|c|c|c|}
\hline
 & \multicolumn{3}{|c|}{Energy in the c.m.s. $\sqrt{s}$, GeV}\\
\cline{2-4}
& 200 & 546 & 900\\
\hline
 $<n>$ & 21.3$\pm$0.80 & 29.1$\pm$0.9 & 34.6$\pm$1.2\\
\hline
 $D$ & 10.9$\pm$0.4 & 16.3$\pm$0.4 & 20.7$\pm$0.6\\
\hline
 $<n>/D^2$ & 0.18$\pm$0.02 & 0.11$\pm$0.01 & 0.08$\pm$0.01\\
\hline
 $k^{-1}$ & 0.22 & 0.27 & 0.31\\
\hline
 $b$ & 0.82 & 0.88 & 0.92\\
\hline
 $<n_c>$ & 2.65 & 3.46 & 4.55\\
\hline
 $D_c$ & 2.77 & 4.11 & 6.01\\
\hline
 $<n_c>/D_c^2$ & 0.34 & 0.20 & 0.12\\
\hline
 $<n>/k$ & 4.65 & 7.88 & 10.31\\
\hline
 $<N>$ & 8.07 & 8.41 & 7.60\\
\hline
\end{tabular}$$
\newpage
 \begin{center}
{\sc Table 2}\\
 Characteristics of the Multiplicity Distributions of Charged Hadrons\\
in ${\it p}Ta$-collisions\\
\end{center}
$$\begin{tabular}{|c|c|c|c|c|}
\hline
 & \multicolumn{4}{|c|}{Energy per nucleon in the lab. system, GeV}\\
\cline{2-5}
 & 2.48 & 4.30 & 5.48 & 9.94\\
\hline
 $<n>$ & 2.98$\pm$0.06 & 4.73$\pm$0.07 & 6.12$\pm$ 0.09 & 7.84$\pm$0.13\\
\hline
 $<n>/D^2$ & 1.38$\pm$0.05 & 0.74$\pm$0.05 & 0.53$\pm$0.05 & 0.30$\pm$0.03\\
\hline
 $k^{-1}$ & -0.08$\pm$0.01 & 0.07$\pm$0.01 & 0.15$\pm$0.01 & 0.31$\pm$0.02\\
\hline
 $b$ & -0.30 & 0.26 & 0.47 & 0.71\\
\hline
 $<n_c>$ & 0.86$\pm$0.04 & 1.18$\pm$0.05 & 1.41$\pm$0.09 & 1.97$\pm$0.12\\
\hline
 $<n>/k$ & -0.27 & 0.34 & 0.91 & 2.40\\
\hline
 $D_c$ & - & 0.45 & 0.82 & 1.71\\
\hline
 $<n_c>/D_c^2$ & - & 5.83 & 2.08 & 1.47\\
\hline
 $<N>$ & 3.47$\pm$0.18 & 4.01$\pm$0.21 & 4.34$\pm$0.22 & 4.22$\pm$0.23\\
\hline
\end{tabular}$$
\bigskip

 \begin{center}
{\sc Table 3}\\
 Characteristics of the Multiplicity Distributions of Charged Hadrons\\
in ${\it d}Ta$-collisions\\
\end{center}
$$\begin{tabular}{|c|c|c|c|}
\hline
 & \multicolumn{3}{|c|}{Energy per nucleon in the lab. system, GeV}\\
\cline{2-4}
 & 2.48 & 4.30 & 5.18\\
\hline
 $<n>$ & 4.46$\pm$0.07 & 7.15$\pm$0.10 & 7.68$\pm$0.21\\
\hline
 $<n>/D^2$ & 0.68$\pm$0.03 & 0.40$\pm$0.05 & 0.31$\pm$0.03\\
\hline
 $k^{-1}$ & 0.09$\pm$0.01 & 0.22$\pm$0.01 & 0.29$\pm$0.03\\
\hline
 $b$ &0.29 & 0.61 & 0.69\\
\hline
 $<n_c>$ & 1.20$\pm$0.06 & 1.65$\pm$0.09 & 1.90$\pm$0.12\\
\hline
 $<n>/k$ & 0.40 & 1.57 & 2.23\\
\hline
 $D_c$ & 0.50 & 1.23 & 1.59\\
\hline
 $<n_c>/D_c^2$ & 4.80 & 1.09 & 0.75\\
\hline
 $<N>$ & 3.72$\pm$0.20 & 4.22$\pm$0.22 & 4.04$\pm$0.25\\
\hline
\end{tabular}$$

\newpage

 \begin{center}
{\sc Table 4}\\
 Characteristics of the Multiplicity Distributions of Charged Hadrons\\
in $HeTa$-collisions\\
\end{center}
$$\begin{tabular}{|c|c|c|c|}
\hline
 & \multicolumn{3}{|c|}{Energy per nucleon in the lab. system, GeV}\\
\cline{2-4}
 & 2.48 & 4.30 & 5.18\\
\hline 
 $<n>$ & 7.64$\pm$0.14 & 10.81$\pm$0.15 & 11.79$\pm$0.25\\
\hline
 $<n>/D^2$ & 0.42$\pm$0.03 & 0.21$\pm$0.02 & 0.16$\pm$0.02\\
\hline
 $k^{-1}$ & 0.22$\pm$0.02 & 0.35$\pm$0.03 & 0.45$\pm$0.05\\
\hline
 $b$ & 0.61 & 0.79 & 0.84\\
\hline
 $<n_c>$ & 1.67$\pm$0.08 & 2.41$\pm$0.10 & 2.88$\pm$0.16\\
\hline
 $<n>/k$ & 1.56 & 3.76 & 5.33\\
\hline
 $D_c$ & 1.22 & 2.38 & 3.12\\
\hline
 $<n_c>/D_c^2$ & 0.89 & 0.42 & 0.29\\
\hline
 $<N>$ & 4.19$\pm$0.14 & 4.48$\pm$0.23 & 4.09$\pm$0.22\\
\hline
\end{tabular}$$
\bigskip

 \begin{center}
{\sc Table 5}\\
 Characteristics of the Multiplicity Distributions of Charged Hadrons\\
in $CTa$-collisions\\
\end{center}
$$\begin{tabular}{|c|c|c|}
\hline
 & \multicolumn{2}{|c|}{Energy per nucleon in the lab. system, GeV}\\
\cline{2-3}
 & 2.48 & 4.30\\
\hline
 $<n>$ & 12.73$\pm$0.61 & 19.75$\pm$0.39\\
\hline
 $<n>/D^2$ & 0.18$\pm$0.02 & 0.08$\pm$0.01\\
\hline
 $k^{-1}$ & 0.36$\pm$0.05 & 0.56$\pm$0.04\\
\hline
 $b$ & 0.82 & 0.92\\
\hline
 $<n_c>$ & 2.66$\pm$0.10 & 4.55$\pm$0.12\\
\hline
 $<n>/k$ & 4.66 & 11.03\\
\hline
 $D_c$ & 2.78 & 6\\
\hline
 $<n_c>/D_c^2$ & 0.35 & 0.12\\
\hline
 $<N>$ & 4.77$\pm$0.22 & 4.29$\pm$0.24\\
\hline
\end{tabular}$$            

\newpage

\section{Figure Captions}

\ \hskip+0.5cm Fig.1  Energy dependence of $R$ and $R_c$ in $pp$-and 
$\bar{p}p$-collisions. + for $R$, $\bullet$  for  $R_c$.

Fig.2  Energy dependence of $R$ in $pTa (\bullet), dTa (\blacktriangledown),
HeTa (\, \vrule height7pt width7pt depth0pt\,),
CTa (\blacktriangleup)$-collisions and $R_c$ in
$pTa (\circ), dTa (\bigtriangledown), HeTa (\Box ), CTa (\bigtriangleup 
)$-collisions.

Fig.3  Energy dependence of the parameter $k^{-1}$ in $A_iTa$- collisions.
$pTa (\bullet)$, $dTa (\circ )$, $HeTa (+)$, $CTa$ $(\bigtriangledown)$.

Fig.4  Energy dependence of $<n>$ in $A_iTa$- collisions,
$pTa (+)$, $dTa$ $(\bigtriangleup)$, $HeTa (\Box )$, $CTa (\bigtriangledown)$.

Energy dependence of $<n_c>$ in $A_iTa$- collisions, 
$pTa (\bullet)$, $dTa (\blacktriangleup)$,\linebreak $HeTa 
(\, \vrule height7pt width7pt depth0pt\,)$, $CTa (\blacktriangledown).$

Fig.5  Dependence $D(<n>)$ in $e^+e^-$, $\bar{p}p$, $pp$, $A_iTa$- collisions,
$e^+e^- (\times)$, $\bar{p}p$, $pp$ $(\bullet)$, $pTa (\bigtriangleup)$,
$dTa$ $(\bigtriangledown)$, $HeTa$ $(\blacktriangledown)$, $CTa$ 
$(\blacktriangleup)$,
curves 1 and 2.

Dependence $D_c(<n_c>)$, in $e^+e^-$, $\bar pp$, $pp$, $A_iTa$-collisions,
$e^+e^- (\circ)$, $\bar{p}p$, $pp (+)$, $pTa 
(\bullet)$,
$dTa (\Box )$, $HeTa (\, \vrule height7pt width7pt depth0pt\,)$, 
$CTa (\sqsubset\hskip-1mm\sqsupset)$, curve 3.

Fig.6  Total multiplicity distribution of charged secondaries in 
$dTa$- collisions at 5.18 AGeV, the curve corresponds to Eq. (1).
\clearpage

\begin{figure}[p]
\begin{center}\mbox{\epsfig{file=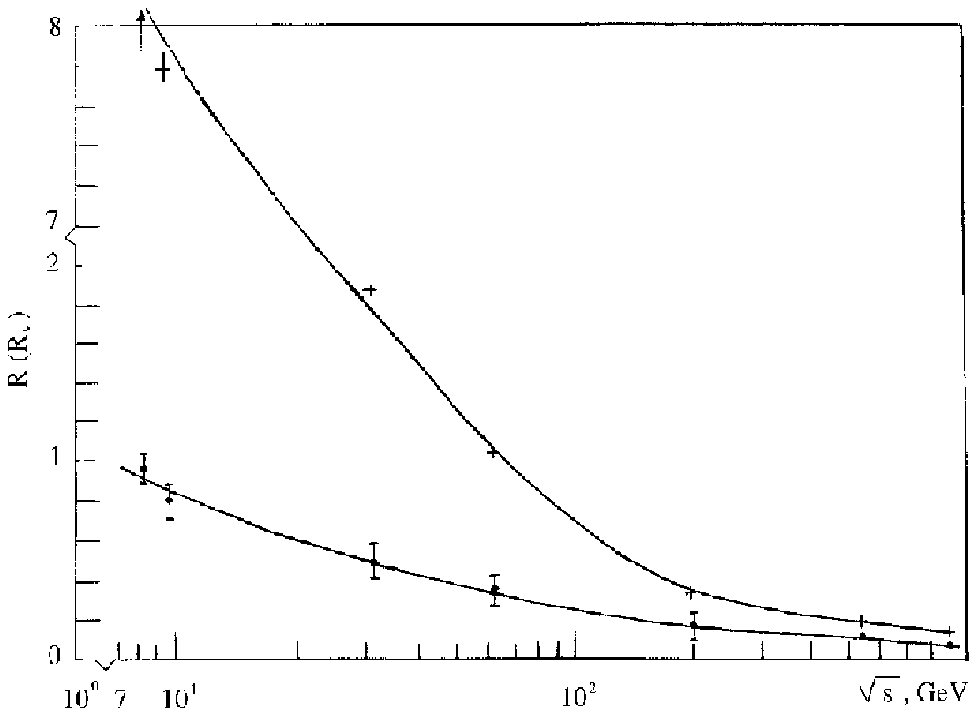,height=10cm}}\end{center}
\caption{}
\end{figure}

\begin{figure}[p]
\begin{center}\mbox{\epsfig{file=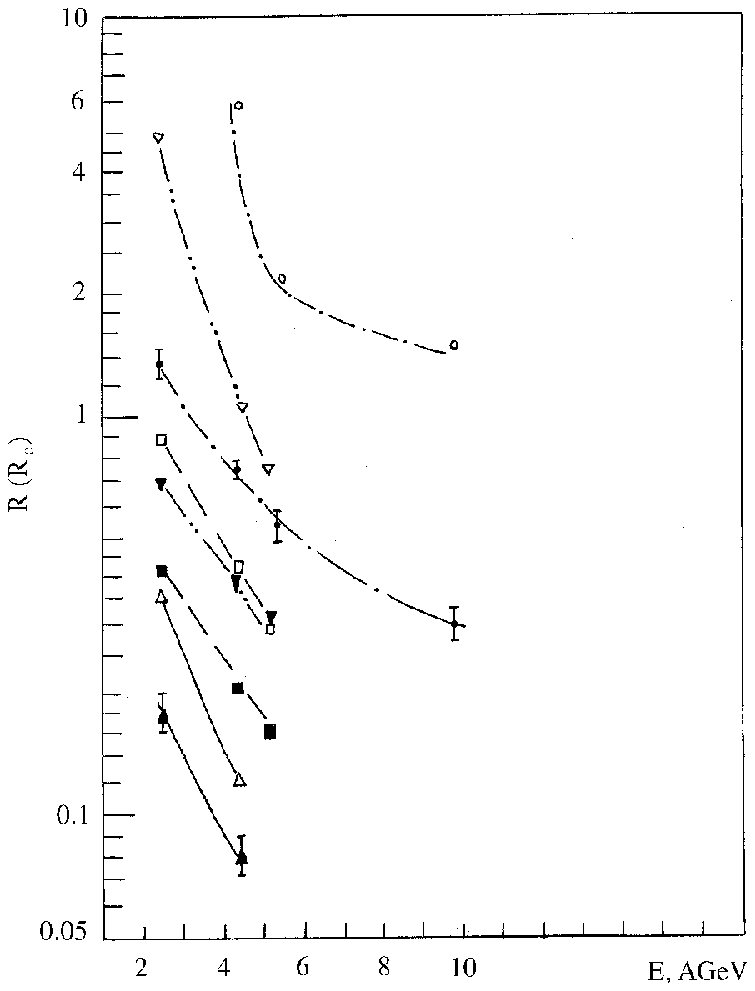,height=10cm}}\end{center}
\caption{}
\end{figure}

\begin{figure}[p]
\begin{center}\mbox{\epsfig{file=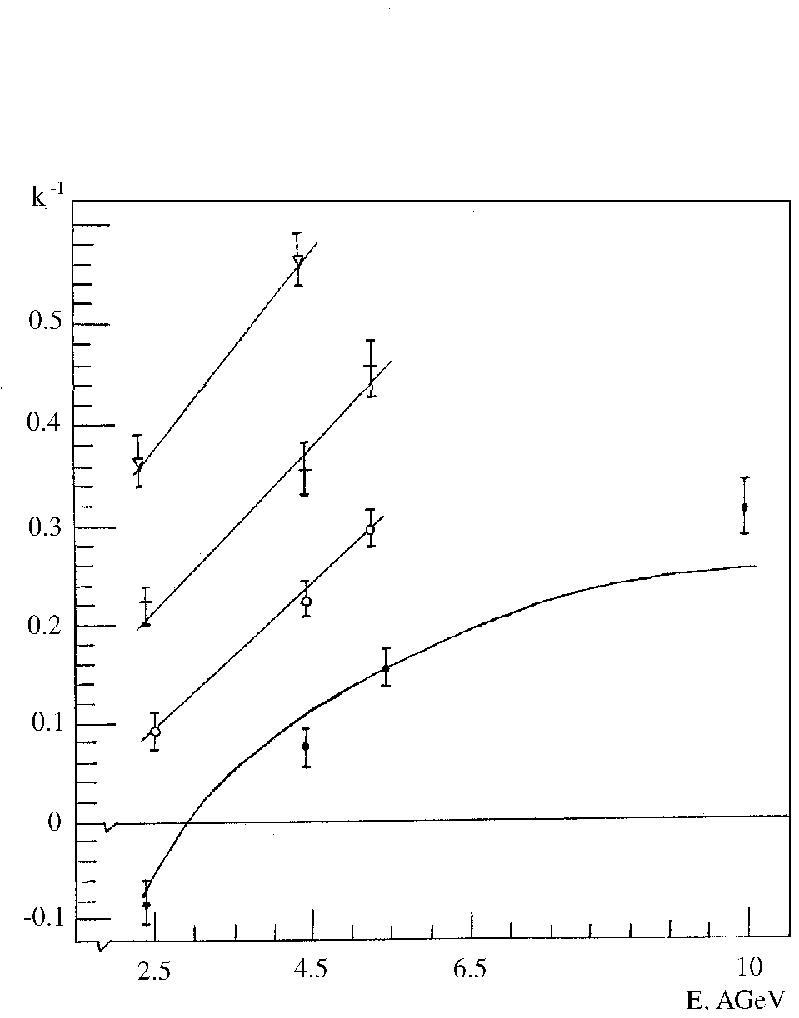,height=10cm}}\end{center}
\caption{}
\end{figure}

\begin{figure}[p]
\begin{center}\mbox{\epsfig{file=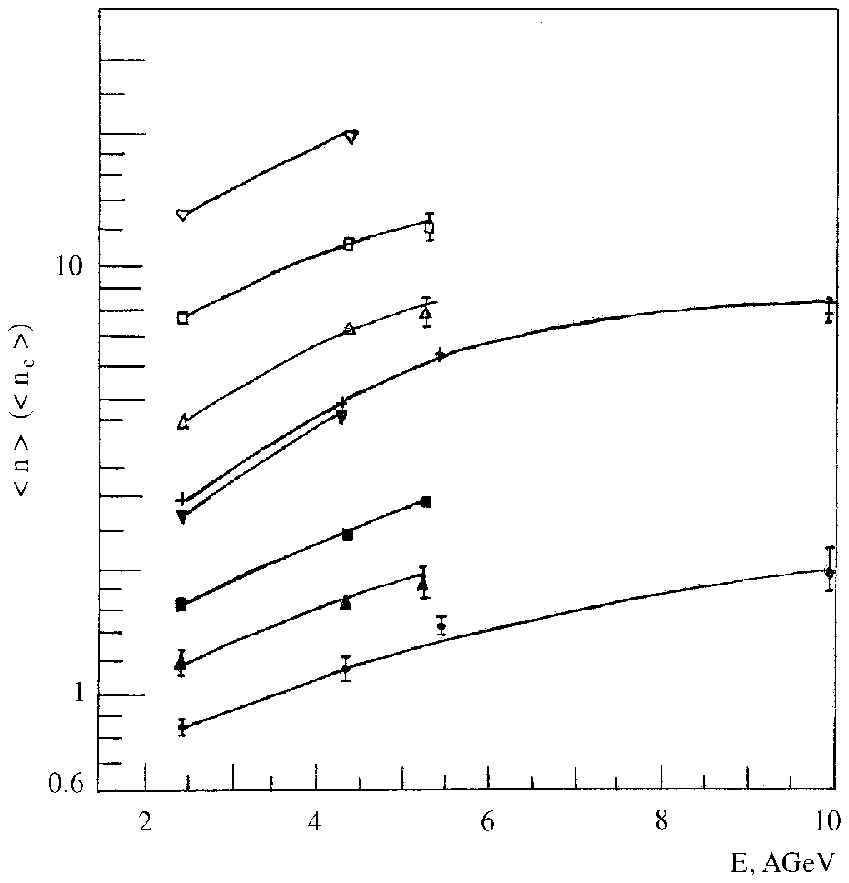,height=10cm}}\end{center}
\caption{}
\end{figure}

\begin{figure}[p]
\begin{center}\mbox{\epsfig{file=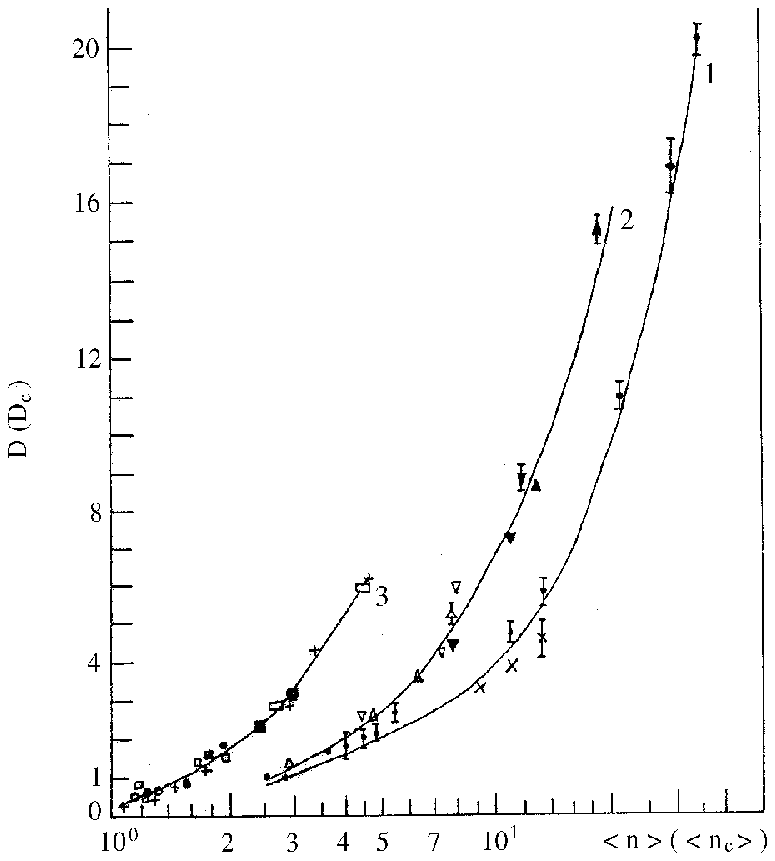,height=10cm}}\end{center}
\caption{}
\end{figure}

\begin{figure}[p]
\begin{center}\mbox{\epsfig{file=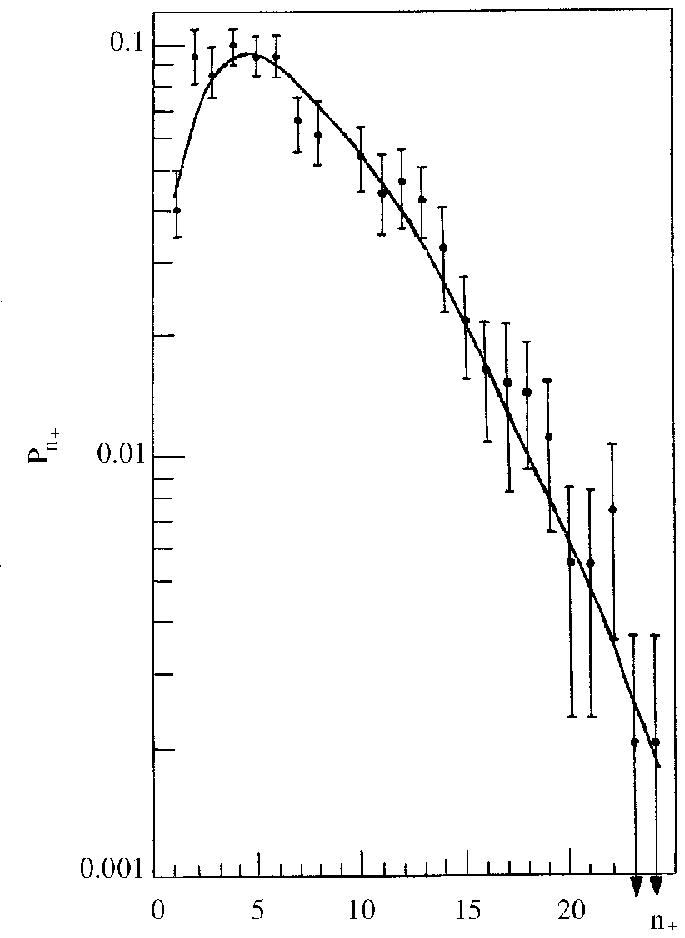,height=10cm}}\end{center}
\caption{}
\end{figure}

\end{document}